\documentclass{aa}

\usepackage{graphicx}
\usepackage{txfonts}
\usepackage{graphics,graphicx}
\usepackage{amsmath}	
\usepackage{amssymb}	
\usepackage{savesym}
\usepackage{enumitem}
\usepackage{mathrsfs,dsfont}
\usepackage{multirow,multicol}
\usepackage{captcont}
\usepackage{float}
\usepackage[export]{adjustbox}
\usepackage{booktabs}
\usepackage{epstopdf}
\usepackage{soul}
\usepackage{xcolor}
\usepackage[colorlinks=true,linkcolor=blue,citecolor=blue,urlcolor=black]{hyperref}
\usepackage{orcidlink} 
\usepackage{cleveref}
\usepackage{verbatim}
\usepackage{makecell}

\def\be{\begin{equation}}
\def\ee{\end{equation}}
\def\kms{{\rm \,km\,s^{-1}}}

\def\Myr{{\rm \,Myr}}
\def\Gyr{{\rm \,Gyr}}

\def\Mpc{{\rm \,Mpc}}
\def\kpc{{\rm \,kpc}}
\def\cm{{\rm \,cm}}

\def\keV{{\rm \,keV}}

\newcommand{\orcid}[1]{\orcidlink{#1}}

\graphicspath{{figures/}}

\begin{document}

\title{AGN Feedback Turbulence in Cool-core Clusters: \\ Evidence from XRISM }


\author{Congyao Zhang\orcid{0000-0001-5888-7052}\thanks{\email{cyzhang@astro.uchicago.edu}}\inst{\ref{aff1},\ref{aff2}}
\and Eugene Churazov\orcid{0000-0002-0322-884X}\inst{\ref{aff3},\ref{aff4}}
\and Irina Zhuravleva\orcid{0000-0001-7630-8085}\inst{\ref{aff2}}
\and Hannah McCall\orcidlink{0000-0003-3537-3491}\inst{\ref{aff2}}
\and Annie Heinrich\orcidlink{0000-0002-7726-4202}\inst{\ref{aff2}}
\and \\ William R. Forman\orcidlink{0000-0002-9478-1682}\inst{\ref{aff5}}
\and Ildar Khabibullin\orcidlink{0000-0003-3701-5882}\inst{\ref{aff7},\ref{aff3}}
\and Michal Zajaček\orcid{0000-0001-6450-1187}\inst{\ref{aff1}}
\and Norbert Werner\orcid{0000-0003-0392-0120}\inst{\ref{aff1}}
}
\institute{
Department of Theoretical Physics and Astrophysics, Faculty of Science, Masaryk University, Brno 61137, Czechia\label{aff1}
\and Department of Astronomy and Astrophysics, The University of Chicago, Chicago, IL 60637, USA\label{aff2}
\and Max Planck Institute for Astrophysics, Karl-Schwarzschild-Str. 1, D-85741 Garching, Germany\label{aff3}
\and Space Research Institute (IKI), Profsoyuznaya 84/32, Moscow 117997, Russia\label{aff4}
\and Harvard-Smithsonian Center for Astrophysics, 60 Garden St., Cambridge, MA, 02138 USA\label{aff5}
\and Rudolf Peierls Centre for Theoretical Physics, Department of Physics, University of Oxford, Clarendon Laboratory, Parks Rd, Oxford, OX1 3PU, United Kingdom\label{aff7}
}

\date{Received xxx; accepted xxx 
}

\abstract{
The XRISM X-ray Observatory has revealed evidence for a ``double-cascade'' scenario in the Perseus cluster, in which active galactic nucleus (AGN) feedback and mergers drive turbulence over distinct radial ranges and spatial scales. Whether this gas kinematic feature is common among cool-core clusters remains unclear. Here, we revisit XRISM observations of A1795 and A2029 to investigate the signatures of AGN feedback and their contribution to the gas velocity fields. In A1795, AGN feedback dominates the velocity dispersion in the core, similar to Perseus, and may provide turbulent heating at a rate comparable to radiative cooling. In contrast, we obtain only an upper limit on the AGN feedback contribution in A2029, where merger-driven motions dominate. Tailored simulations indicate that an off-axis minor merger $\simeq1.6\Gyr$ ago displaced the central low-entropy gas in A2029, potentially suppressing AGN activity. This gas is now returning to the center, possibly facilitating the re-establishment of AGN feedback. We propose an evolutionary view of the double-cascade scenario, in which merger-driven motions dominate during and shortly after significant mergers, while feedback-driven turbulence re-emerges in cluster cores as systems relax. This framework provides a basis for interpreting XRISM kinematic measurements of (weak) cool-core clusters. Our findings highlight the importance of spatially resolved gas kinematics and detailed modeling for identifying double-cascade features and disentangling the contributions of AGN feedback and mergers.
}


   \maketitle
%

\section{Introduction}

Galaxy clusters are assembled through merger and accretion of smaller structures from the cosmic web in the hierarchical structure formation scenario (see \citealt{Kravtsov2012,Vikhlinin2014} for reviews). Both major mergers and significant minor mergers can substantially redistribute the cluster mass and energy content \citep[e.g.,][]{Poole2006,Paul2011}, in particular the baryonic component. They drive prominent shocks and strong turbulence during the first $\sim1\Gyr$, corresponding to their most violent phase. The merger remnants then undergo a relatively gentle evolution over the subsequent few Gyr until the next significant merger event. During this phase, the clusters relax and develop cool cores in their atmospheres through radiative cooling. Mergers can still have long-term impacts at this stage, often in the form of gas sloshing \citep[e.g.,][]{Ascasibar2006}, which gradually dissipates gravitational potential energy. Sloshing may partially offset radiative cooling losses at large radii \citep{ZuHone2010,XRISM2026_Perseus}, but it is often insufficient in the very central region, where the gas density is high and the cooling timescale is short.

Radio-mode active galactic nucleus (AGN) feedback provides a promising heating solution within the cool cores \citep{Fabian2012,Werner2019}. Energy released by the central supermassive black hole is injected into the surrounding medium through jets and inflated bubbles \citep{Churazov2001,Churazov2002}, which further drive shocks, sound waves, gravity waves, and turbulence \citep[e.g.,][]{Fabian2006,Zhuravleva2014,Zhang2018}. These processes leave observable imprints on both the thermal and nonthermal structures of the intracluster medium (ICM), many of which have been revealed by high-spatial-resolution X-ray and radio observations (e.g., Chandra, XMM-Newton, and LOFAR).

The recently launched XRISM X-ray Observatory provides unprecedented kinematic information on the hot atmospheres of galaxy clusters \citep{Tashiro2021}. Spatially mapping line shifts and line broadening allows us to directly characterize the kinetic energy injection, transport, and turbulent cascade associated with both AGN feedback and mergers \citep[e.g.,][]{XRISM2025_Centaurus,XRISM2026_Perseus,XRISM2026_Virgo}. 
From a theoretical point of view, mergers inject energy on large scales, of order hundreds of kpc or larger, comparable to the characteristic size of the infalling substructures and their orbital trajectories, whereas AGN feedback injects energy on much smaller scales and is more spatially concentrated in cluster cores. XRISM has provided the first observational evidence for this picture by revealing a significant enhancement of the ICM kinetic energy on small scales within a projected radius of $\simeq60\kpc$ in the Perseus cluster (\citealt{XRISM2026_Perseus}; see also Fig.~\ref{fig:perseus_model}). This finding implies that, in the Perseus cluster, AGN feedback efficiently drives turbulence within the cool core. \citet{XRISM2026_Perseus} proposed the ``double-cascade'' scenario, in which feedback and mergers drive turbulence predominantly at distinct spatial scales and different radii.

A natural follow-up question is whether this double-cascade picture is universal among cool-core clusters. Answering this question will provide critical insight into how AGN feedback distributes its energy and how efficiently that energy couples to the inner cluster atmosphere. Dozens of galaxy clusters, including several cool-core systems, have been observed by XRISM. However, isolating the contribution of AGN feedback to the observed velocity dispersion remains challenging because sloshing motions often coexist with feedback-driven structures in cluster cores. Perseus is a unique target for XRISM/Resolve due to its low redshift ($z\simeq0.017$) and high X-ray surface brightness \citep[e.g.,][]{Edge1990}, which together allow the AGN-feedback-dominated region to be spatially resolved and the velocity measurements to extend significantly beyond it. The two dominant drivers can therefore be directly separated based on a simple scale argument \citep{XRISM2026_Perseus}. Few other targets satisfy such conditions. Detailed analysis of their merger history is thus essential for characterizing the contribution of AGN feedback. For this reason, well-relaxed clusters that have not experienced significant mergers within the past few Gyr, or systems undergoing a single merger with a well-constrained configuration, are ideal targets.

A1795 and A2029 are among the most suitable systems in the current XRISM cluster sample for probing the kinematic signatures of AGN feedback. Both are well-known cool-core clusters with extensive multi-wavelength observations and a long history of X-ray studies. Their velocity radial profiles have been measured using multiple XRISM pointings \citep{Sarkar2026,XRISM2025_A2029_Outer}, enabling the characterization of merger-induced gas motions on large scales. In this work, we revisit the XRISM measurements of these systems by carefully modeling and separating the contributions from AGN feedback and merger-driven motions. We find that AGN feedback makes a dominant contribution to the velocity dispersion radial profile within $\simeq100\kpc$ in A1795, indicating a kinematic sphere of influence similar to that found in Perseus \citep{XRISM2026_Perseus}. Despite large uncertainties, the feedback-associated heating rate of turbulence is comparable to the radiative cooling rate in both A1795 and Perseus, highlighting the significant role of feedback-driven turbulence in preventing catastrophic overcooling of the ICM. In contrast, the AGN feedback contribution is subdominant in A2029, suggesting that the cluster is in a transitional stage in which AGN activity has been re-established following recent merger-induced quenching, while merger-driven motions still make a substantial contribution to the observed kinematics. This interpretation is broadly consistent with the lack of cold gas despite the presence of newly launched radio jets with an estimated age of $\sim100\Myr$ in A2029 \citep{Timmerman2022,Olivares2023}. Based on these results, we propose a unified evolutionary picture for the interplay between mergers and AGN feedback, providing a framework for interpreting the kinematic properties of a broader sample of XRISM-observed clusters, such as A3571 \citep{Aihara2026,McCall2026}, Centaurus \citep{XRISM2025_Centaurus}, and Coma \citep{XRISM2025_Coma,Gatuzz2026}. We emphasize that the moderate velocity dispersions commonly found in the growing sample of XRISM-observed clusters ($\simeq100$–$200\kms$) can arise from complex and diverse dynamical processes. Detailed case-by-case modeling is therefore essential to interpret these measurements and disentangle the contributions of mergers and AGN feedback.

This paper is organized as follows. Section~\ref{sec:model} introduces the double-cascade scenario and its modeling using Perseus as an example. In Section~\ref{sec:result}, we extend the analysis to A1795 and A2029. Section~\ref{sec:discussions} discusses the physical implications of our findings, and Section~\ref{sec:conclusions} summarizes our main conclusions. Throughout the paper, we assume a flat cosmology with $H_0=70\kms\Mpc^{-1}$ and $\Omega_{\rm m}=0.30$, under which $1' \simeq70\kpc$ and $90\kpc$ at the A1795 and A2029's redshift, respectively.

\section{Modeling the double-driver scenario: Perseus as an example} \label{sec:model}

The multiple XRISM Perseus pointings enable gas motions to be resolved over a broad range of scales ($\sim50-250\kpc$), benefiting from the cluster's stratified X-ray surface brightness distribution for the following reason.

The velocity dispersion measured along a given line of sight (LOS) reflects motions on (or below) the gas core scale that contributes the bulk of the X-ray flux in projection -- the so-called effective length scale $\ell_{\rm eff}$ \citep{Zhuravleva2012}. Accounting for this effective length is essential for interpreting XRISM velocity measurements \citep[e.g.,][]{XRISM2026_Perseus,Zhang2026,XRISM2026_Virgo}, especially when comparing X-ray-weighted velocities across different clusters or within the same cluster at different radii. In Perseus (and also the other cool-core clusters discussed in this paper), the radial profile of $\ell_{\rm eff}$ is a monotonically increasing function of radius, with a flattening in the very inner dense core. The comparable velocity dispersions ($\simeq170\kms$) measured in the central and outer regions of Perseus probe spatial scales differing by a factor of $\simeq5$, which cannot be explained by a single velocity cascade. At least two significant drivers are therefore required: one concentrated in the inner core and naturally associated with AGN feedback, and another operating on much larger scales associated with the merger. We generally expect the former component to contribute little to the large-scale bulk velocity and its velocity dispersion to decline with radius, while the latter drives an increase in velocity dispersion with radius on scales larger than the atmospheric scale height and can form significant bulk motions.

\begin{figure}
\centering
\includegraphics[width=0.95\linewidth]{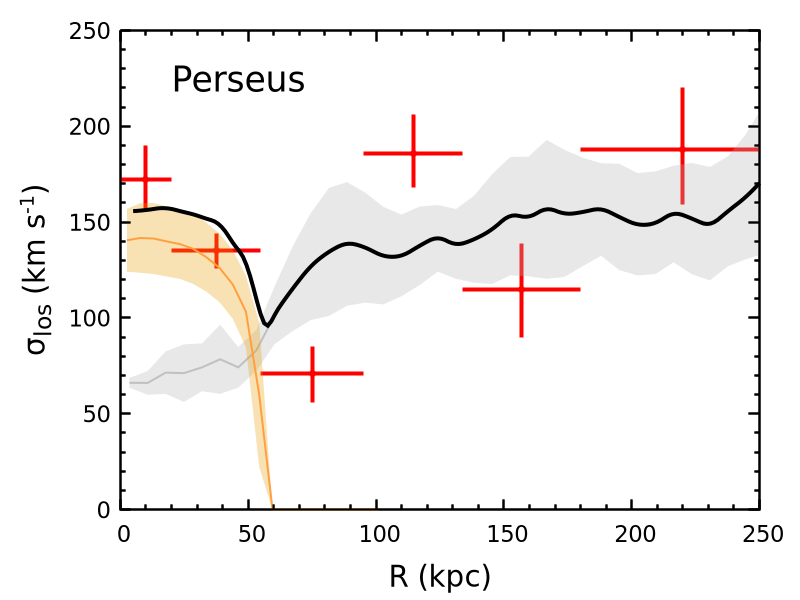}
\vspace{-5pt}
\caption{Radial profile of the velocity dispersion (red points) in Perseus along the NW direction, measured from the XRISM Performance Verification (PV) observations. The measurements indicate the presence of two dominant kinematic drivers \citep{XRISM2026_Perseus}: one powered by AGN feedback and the other associated with mergers. The corresponding theoretical models are shown by the light-orange and grey lines, respectively. The AGN feedback component is modeled as a uniform Gaussian random velocity field with $\sigma_{\rm agn}=150\kms$, $\ell_{\rm inj}=20\kpc$ and $r_{\rm agn}=55\kpc$ (see Section~\ref{sec:model} for more details). The shaded region indicates the standard deviation of the velocity dispersion in each radial bin from the turbulent simulation. The thick black line shows the total model, obtained by adding the two components in quadrature, which provides a good description of the XRISM measurements. }
\label{fig:perseus_model}
\end{figure}

In Fig.~\ref{fig:perseus_model}, we model the contributions from these two drivers to the velocity dispersion radial profile in Perseus, shown as the grey and light-orange shaded regions, respectively. The merger-driven component is adopted from the turbulent simulations presented in \citet{XRISM2026_Perseus}, where large-scale turbulence is driven in a static Perseus-like gravitational potential with an energy injection scale of $\simeq500\kpc$ (see Appendix~\ref{sec:appendix:sim} for more details). We extract the radial profile from a sector with an opening angle of $90^\circ$ for comparison with the observational measurement along the northwestern (NW) arm. The scatter corresponds to the standard deviation of the velocity dispersion and reflects its azimuthal spatial variation within the sector. This model closely resembles the merger simulation results tailored to Perseus \citep{Bellomi2026}. Note that Perseus is not yet fully relaxed. It exhibits complex sloshing cold-front structures ranging from the inner spiral-like features \citep[e.g.,][]{Fabian2000,Churazov2003} to the giant cold front at a radius of $\sim700\kpc$ \citep{Simionescu2012,Walker2018}, which may correspond to different past merger events \citep{Zhang2026}. It is therefore non-trivial to reproduce these sloshing structures numerically in detail (see further discussions in \citealt{Bellomi2024,Zhang2026}), which may complicate the separation of feedback and merger contributions near the cold front.

Modeling the AGN feedback contribution is substantially more challenging. The injection of kinetic energy is sensitive to the adopted sub-grid prescriptions in numerical simulations (see \citealt{Bourne2023} for a review). \citet{XRISM2025_Simulation} suggested that some recent cosmological simulations may overproduce turbulence in cluster cores due to their AGN feedback implementation. In addition, the long-term interaction between buoyantly rising AGN bubbles, after detaching from the central supermassive black hole, and the ambient ICM remains difficult to capture numerically. These rising bubbles are expected to play an essential role in uplifting low-entropy gas, driving turbulent wakes, subsequently stirring the cluster inner atmosphere \citep{Churazov2000,Churazov2002}. The characteristic scale of these motions is likely comparable to the bubble size, ranging from several to tens of kpc  \citep[e.g.,][]{Li2020,Zhang2022}.

Here we adopt a simple, ad hoc model, assuming that volume-filling turbulence has developed and is uniformly distributed within a given radius $r_{\rm agn}$ (see \citealt{Bellomi2026} for more sophisticated jet simulations in Perseus). In Perseus, this assumption is motivated by the sudden drop in velocity dispersion immediately outside the cold front at $\simeq55\kpc$ in the NW direction. This provides an indication that feedback-driven turbulence is at least partially confined within the sloshing cold front. A NW bubble is flattened and located near the sloshing interface, implying efficient generation and distribution of turbulent motions in the surrounding region. This is also consistent with the morphology of the H$\alpha$ filaments and their LOS velocities in the bubble downstream \citep{Fabian2003,Li2020}.

We model the feedback-driven velocities by generating Gaussian random velocity fields with an underlying Kolmogorov-type energy spectrum, characterized by two free parameters: the injection scale $\ell_{\rm inj}$ and the normalization (equivalent to the 1D turbulent velocity dispersion $\sigma_{\rm agn}$ over the volume; see \citealt{Zhang2026} for more technical details). Unless giant radio lobes on $\sim100\kpc$ scales are generated directly from the cluster center, as in Hydra A or MS0735 \citep[e.g.,][]{Wise2007,McNamara2009}, the feedback-driven turbulence is expected to have a much smaller injection scale than merger-driven turbulence. In Fig.~\ref{fig:perseus_model}, the light-orange line shows the X-ray-weighted velocity dispersion radial profile with $\ell_{\rm inj}=20\kpc$, $\sigma_{\rm agn}=150\kms$, and $r_{\rm agn}=55\kpc$. The scatter is the averaged standard deviation over 100 realizations. The model is insensitive to $\ell_{\rm inj}$ as long as it is comparable to or smaller than the local effective length ($\sim50\kpc$). Note that gravity regulates radial gas motions on scales larger than the Ozmidov length, $\ell_{\rm BV}=\sqrt{3}\sigma_{\rm agn}/N_{\rm BV}$, where $N_{\rm BV}$ is the Brunt–V\"{a}ais\"{a}l\"{a} frequency. Namely, eddies larger than $\ell_{\rm BV}$ are generally suppressed in the bubble wake as the bubbles buoyantly rise, i.e., $\ell_{\rm inj}\lesssim\ell_{\rm BV}$ \citep{Zhang2018}. In Perseus, $N_{\rm BV}$ ranges $\simeq5-15\,\rm{rad\Gyr^{-1}}$ with a peak at $r\simeq50\kpc$ \citep{Churazov2003}. This corresponds to $\ell_{\rm BV}\sim20-40\kpc$ over the radius of $\simeq20-150\kpc$ for the measured velocity amplitude.

\begin{figure}
\centering
\includegraphics[width=0.95\linewidth]{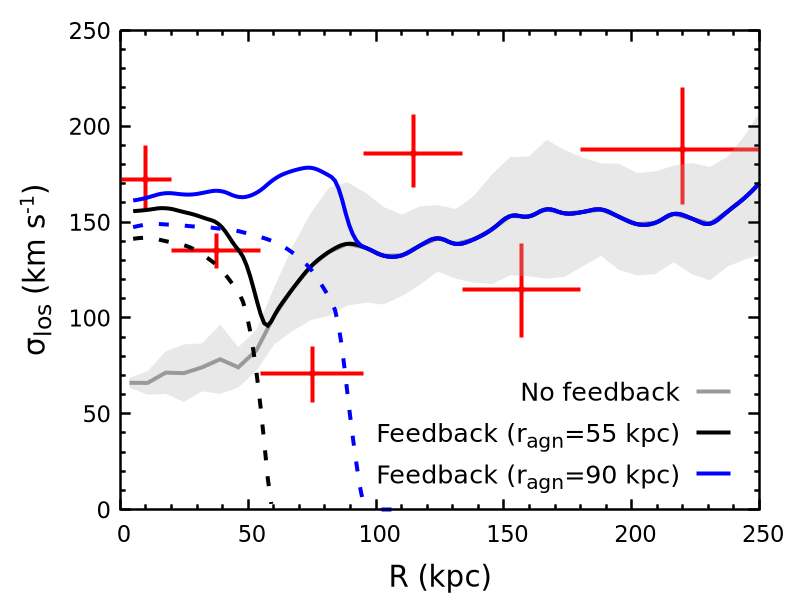}
\vspace{-5pt}
\caption{Predicted radial profiles of the velocity dispersion in Perseus for different extents of the AGN feedback region, $r_{\rm agn}=55\kpc$ (black, same as in Fig.~\ref{fig:perseus_model}) and $r_{\rm agn}=90\kpc$ (blue). The dashed and solid lines show the feedback component and the total model, respectively. The red points reproduce the Perseus data points shown in Fig.~\ref{fig:perseus_model}. The double-driver scenario can produce a relatively flat velocity dispersion radial profile (see Section~\ref{sec:model}).  }
\label{fig:perseus_agn_model}
\end{figure}

To first order, the total velocity dispersion is given by the quadratic sum of the two components shown as the black thick line, consistent with the XRISM measurements. The normalization $\sigma_{\rm agn}$ is the only parameter tuned to visually match our model to the observations. For isotropic Kolmogorov turbulence, the feedback-powered turbulent heating rate scales as $q_{\rm heat}\propto\sigma_{\rm agn}^3/\ell_{\rm inj}$. Any spatial non-uniformity or intermittency in the velocity field may introduce variations in the inferred $\sigma_{\rm los}$ radial profile, which cannot be resolved with current XRISM measurements. The constrained $\sigma_{\rm agn}$ therefore represents only an averaged value over the full volume (see Section~\ref{sec:discussion:heating} for further discussion).

We also emphasize that azimuthal variations in the impact of AGN feedback are expected in galaxy clusters, where many factors can play a role, including but not limited to jet precession, bubble rising trajectories, and sloshing motions. In particular, the feedback-dominated region may be more extended in directions without prominent sloshing cold fronts (probably the situation in A1795; see Section~\ref{sec:result:a1795}). In Fig.~\ref{fig:perseus_agn_model}, we show predictions for the radial profile of $\sigma_{\rm los}$ for different extents of AGN feedback. A `V'-shaped radial profile is not a necessary signature of the presence of two drivers, although it can reveal an unambiguous transition radius as observed in the NW region of Perseus \citep{XRISM2026_Perseus}. \citet{Heinrich2026} recently found that the impact of AGN feedback extends farther in the southern than in the NW region of Perseus.

\section{Feedback-driven turbulence in A1795 and A2029} \label{sec:result}

\subsection{A1795} \label{sec:result:a1795}

\begin{figure}
\centering
\includegraphics[width=0.95\linewidth]{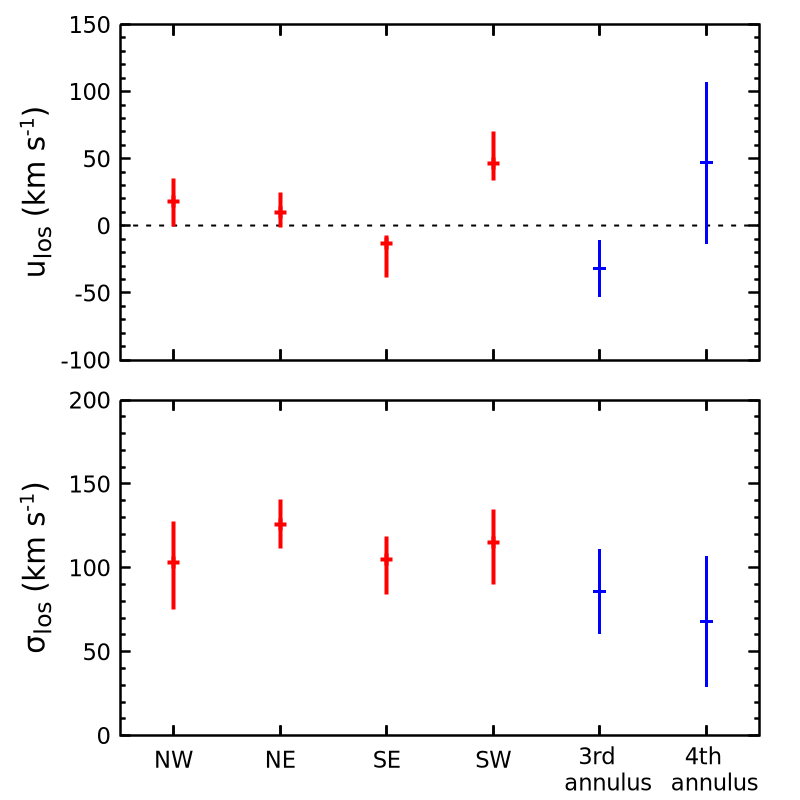}
\vspace{-5pt}
\caption{Bulk velocity (top) and velocity dispersion (bottom) measured across the four quadrant subregions of the central XRISM/Resolve pointing of A1795 (red points). The bulk velocities are measured relative to the BCG ($z_{\rm BCG}=0.06300$). For comparison, the blue points show measurements from the two outermost annuli in the off-axis pointing reported by \citet{Sarkar2026}. No significant bulk motions are detected in the observed regions, suggesting that AGN feedback on small scales makes a substantial contribution to the velocity field in the cluster core (see Section~\ref{sec:result:a1795}). }
\label{fig:a1795_params}
\end{figure}

A1795 is a well-known strong cool-core cluster \citep[e.g.,][]{Fabian2001,Ehlert2015}. It appears dynamically relaxed on large scales ($\gtrsim100\kpc$) and exhibits a remarkably smooth gas distribution in its outskirts \citep{Kovacs2023}. On the other hand, A1795 hosts rich small-scale X-ray and radio structures near its center ($\lesssim80\kpc$; see, e.g., \citealt{Markevitch2001,Giacintucci2014,McCall2025,vanWeeren2026}), likely associated with episodic AGN activity \citep{Kokotanekov2018}.
As a XRISM General Observer (GO) target, the radial velocity profile of A1795 was measured to the north out to $\simeq300\kpc$ \citep{Sarkar2026}. No significant gas bulk motions with respect to the brightest cluster galaxy (BCG) were detected -- the measured velocities are consistent with zero within $2\sigma$ uncertainties. This implies that either the significant large-scale motions are predominantly confined to the plane of the sky, or the gas velocity field is dominated by small-scale motions -- smaller than the local effective length.

We argue that the former scenario is unlikely. In relaxed clusters, large-scale motions are generally attributed to merger-driven sloshing processes (see \citealt{Zuhone2016} for a review), which leave characteristic cold-front imprints and form spiral patterns when the sightline is closely perpendicular to the merger plane. In A1795, only one cold front has been confirmed at a radius of $\simeq80\kpc$ from the cluster center, and it is relatively small in extent in the plane of the sky ($\simeq70\kpc$; see \citealt{Markevitch2001,Ghizzardi2010}), despite the cluster having one of the deepest Chandra accumulated exposures among all cluster targets as a calibration source. This suggests that no significant merger event has occurred recently (at least within the past $\sim3$--$5\Gyr$; see, e.g., \citealt{Ascasibar2006}), which is generally consistent with the smooth and symmetric X-ray morphology of the cluster, although residual motions from previous significant mergers may still remain but should have dissipated most of their kinetic energy. Indeed, the Einstein Probe Follow-up X-ray Telescope recently detected weak X-ray surface brightness excesses at radii of $\gtrsim500\kpc$ in the cluster \citep{Tang2026}, which may be associated with such dynamical activity. \citet{Bellomi2024} measured a radial propagation speed of the sloshing arm in the Perseus cluster of $\sim80\kpc\Gyr^{-1}$. This rate may also be applicable to A1795, as the two clusters have similar radial profiles of the Brunt–V\"{a}ais\"{a}l\"{a} frequency. If so, it would imply that the merger occurred $\gtrsim6\Gyr$ ago.

We note that velocity radial profiles may partially obscure the dipole-like bulk velocity pattern often expected in sloshing clusters. To characterize the azimuthal variation of the velocity field, we re-analyze the central XRISM pointing of A1795 (ObsID~201087010) by dividing the XRISM/Resolve field of view (FOV) into four quadrant subregions ($1.5'\times1.5'$). The data reduction and spectral modeling follow \citet{Sarkar2026}, using the latest calibration database (CalDB version 20260315). The spectra are modeled with a single-temperature collisional-ionization-equilibrium component and spatial-spectral-mixing (SSM) correction, without including the off-axis pointing. Fig.~\ref{fig:a1795_params} shows our measurements, confirming the absence of strong bulk motions in the core of A1795, i.e., $|u_{\rm los}|\lesssim50\kms$. The velocity dispersions are higher than bulk velocities, but their spatial variations are small, indicating that the velocity field is dominated by small-scale motions.

\begin{figure}
\centering
\includegraphics[width=0.95\linewidth]{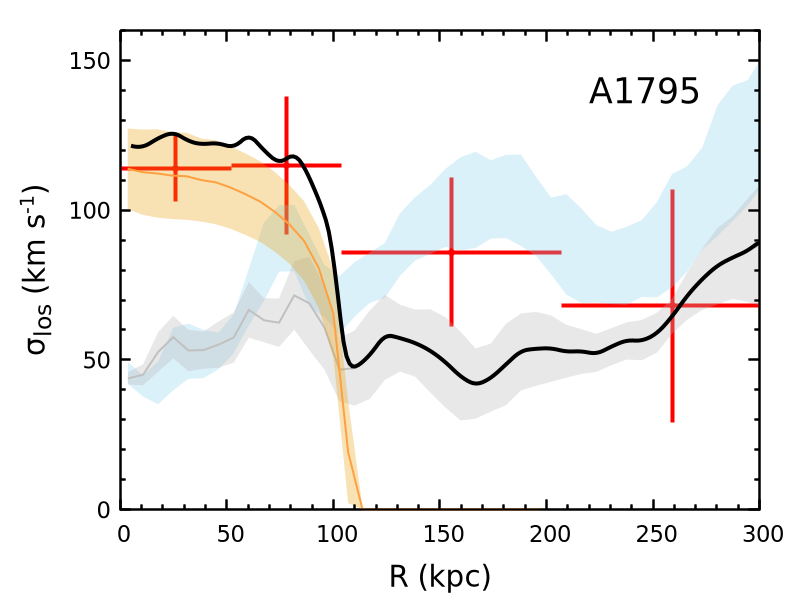}
\vspace{-5pt}
\caption{Same as Fig.~\ref{fig:perseus_model} but for A1795. The red points show the XRISM measurements \citep{Sarkar2026}. The AGN feedback component (light-orange) is modeled with $\sigma_{\rm agn}=120\kms$, $\ell_{\rm inj}=20\kpc$, and $r_{\rm agn}=105\kpc$. The merger component (grey; baseline model) is modeled using the same turbulent simulation as for Perseus. The profile extracted from an earlier snapshot of the simulation (blue) is shown to illustrate the potential systematic uncertainty in the merger contribution (see also Fig.~\ref{fig:turb_sim_perseus}). In either case, AGN feedback dominates the velocity dispersion in the core of A1795 (see Section~\ref{sec:result:a1795}).  }
\label{fig:a1795_model}
\end{figure}

All of this observational evidence indicates that AGN feedback makes a substantial contribution to the velocity field in the cluster core. Fig.~\ref{fig:a1795_model} shows our double-driver decomposition of the velocity dispersion radial profile of A1795, separating the feedback contribution from the remaining merger-induced motions. The weak merger component is modeled using the decaying turbulence from our Perseus simulation to represent the pre-existing turbulent motions in the background (\citealt{XRISM2026_Perseus}; see also Fig.~\ref{fig:turb_sim_perseus}). We note that A1795 and Perseus exhibit similar cool-core structures (as well as the radial profile of $\ell_{\rm eff}$) and comparable gravitational potentials, which are expected to produce similar impacts on the late-stage evolution of decaying turbulence. We use the simulation to capture the stratified turbulence and the residual weak oscillations of the gas core long after the perturbation, which may be related to the cold front found in A1795 \citep{Markevitch2001}. We adopt the snapshot in which the resulting velocity dispersion matches the measurement near $\sim300\kpc$ -- a radius far beyond the cooling radius of the system (grey line, the baseline model hereafter). For comparison, the profile from an earlier snapshot ($\simeq2\Gyr$ earlier) that predicts higher velocity dispersion outside $100\kpc$ is shown as the blue line. It remains consistent with the measurements, but lies toward the upper end of the statistical uncertainty. In either case, the merger-driven component contributes $\lesssim20\%$ of the total kinetic energy within $\simeq70\kpc$. Better measurements of the outer part of the cluster, through either longer exposures or broader azimuthal coverage, would provide tighter constraints on this component in future, but would not significantly affect our conclusions.

The feedback contribution is modeled following the approach described in Section~\ref{sec:model}. The approximately flat velocity dispersion profile out to $\simeq100\kpc$ requires $r_{\rm agn}\gtrsim70\kpc$, providing evidence for the highly extended impact of AGN feedback. Although a sharp cutoff radius cannot be justified in the absence of prominent sloshing cold fronts, this model provides insight into the kinematic sphere of influence of AGN feedback in the system. The model with $r_{\rm agn}=105\kpc$ tends to underestimate the velocity dispersion between $100-200\kpc$ when adopting the baseline merger model, suggesting that the influence of AGN feedback may extend to even larger radii. It may also help explain the relatively uniform velocity dispersion across the core. However, better constraints on the merger contribution are required to confirm this interpretation. Nevertheless, A1795 exhibits prominent signatures of the double-driver scenario, making it the second clear example of this feature after Perseus. The AGN feedback component dominates the inner velocity dispersion field with the scale $\lesssim\ell_{\rm eff}$ (locally $\simeq50-100\kpc$). This is broadly consistent with the rich signatures of AGN feedback revealed by Chandra observations \citep{Kokotanekov2018}. Besides, the flat velocity-dispersion profile disfavors sound waves as the dominant source of the velocity field in A1795, as their contribution is expected to attenuate rapidly with radius. A similar argument has been applied to the XRISM measurements of Perseus and Virgo \citep{XRISM2026_Perseus,XRISM2026_Virgo}.

\begin{figure*}[t]
\centering
\includegraphics[width=0.85\linewidth]{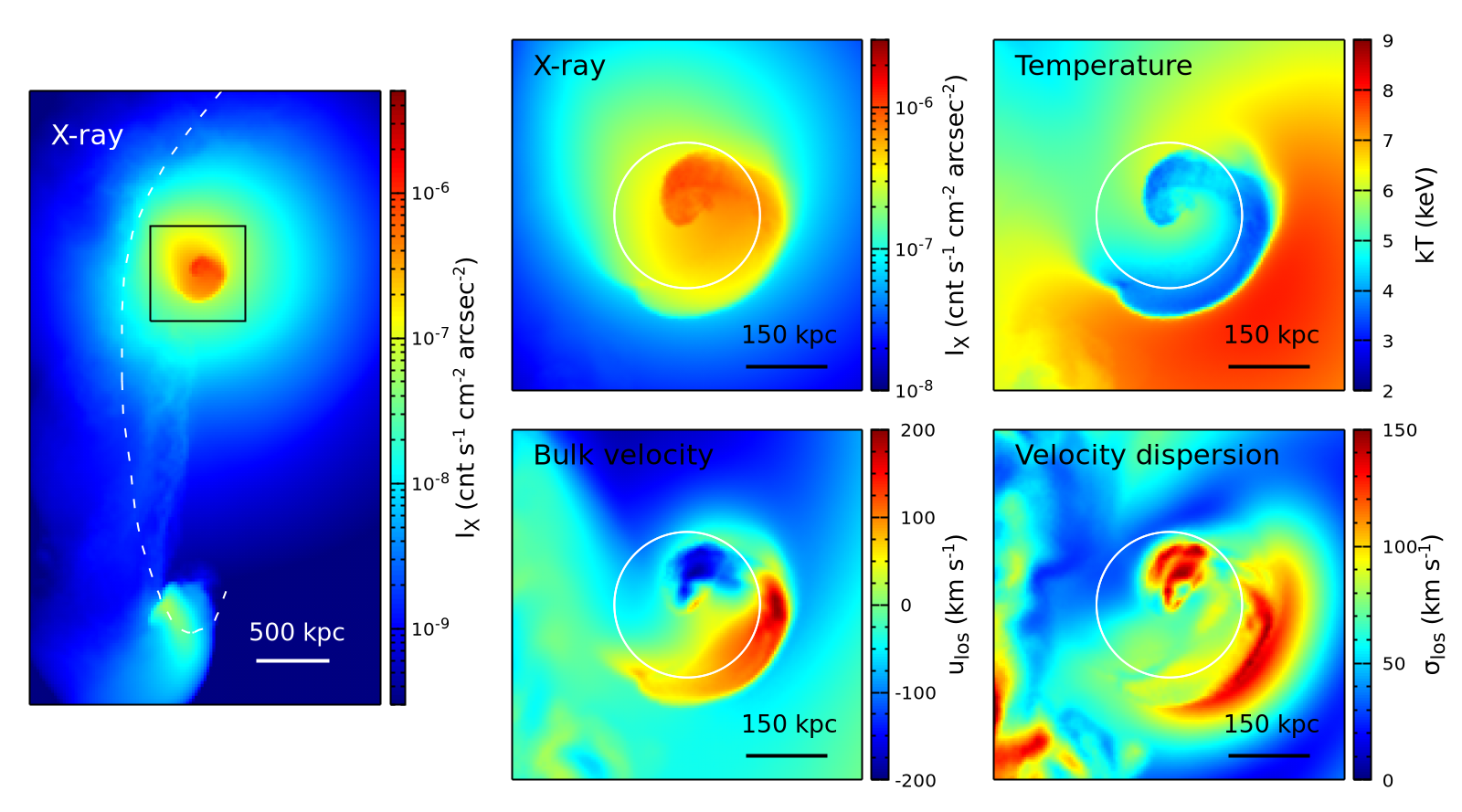}
\vspace{-5pt}
\caption{\textit{Left panel:} Simulated X-ray surface brightness distribution at $t\simeq1.6\Gyr$, which resembles the relative positions of A2029 (the main cluster) and a southern infalling group \citep[see][]{Sohn2019,Mirakhor2022}. The white dashed line indicates the group's infalling trajectory. The group is approaching its primary apocenter. \textit{Right panels:} Zoomed-in maps of the X-ray surface brightness, X-ray-weighted temperature, X-ray-weighted LOS bulk velocity, and velocity dispersion in the region indicated by the black box in the left panel. The white circle approximately indicates the region covered by the XRISM central PV pointing. Our simulation broadly captures key X-ray features in A2029 (see Section~\ref{sec:result:a2029}).  }
\label{fig:merger_mock_a2029}
\end{figure*}

The $\simeq45\kpc$-long X-ray filament extending southward from the BCG is one of the most striking cool-gas structures in A1795 \citep{Fabian2001}, spatially coincident with both H$\alpha$ and molecular filaments \citep[e.g.,][]{Crawford2005,Russell2017,Tamhane2023}. \citet{Fabian2001} proposed a ``cooling wake'' scenario for its origin, which is, however, disfavoured by recent XRISM and LOFAR/MeerKAT observations \citep{Sarkar2026,vanWeeren2026}. The lack of significant relative motion between the BCG and the ICM, together with the close spatial association between the filament and AGN activity revealed in radio observations, instead supports a bubble-driven scenario (\citealt{Russell2017} and see also \citealt{Churazov2001,McNamara2016,Zhang2022}). Buoyantly rising bubbles inflated by the central AGN are able to uplift a significant amount of low-entropy gas from the cluster center and generate bubble-sized turbulent eddies in their wake \citep{Zhang2022}. In this picture, the edge of a ghost bubble provides an alternative interpretation of the cold front identified south of the filaments \citep{Markevitch2001}. The morphology of the filaments, particularly the ``hook'' structure \citep{Ehlert2015}, recalls the well-known ``horseshoe-shaped'' H$\alpha$ filaments in Perseus \citep{Fabian2003,Gendron-Marsolais2018}. This interpretation further supports the presence of an AGN-feedback-dominated region extending over a scale of $\sim70-100\kpc$, and suggests that rising (ghost) bubbles may play a significant role in driving the motions (see Section~\ref{sec:discussions} for further discussion).

\subsection{A2029} \label{sec:result:a2029}

A2029 is a XRISM PV target \citep{XRISM2025_A2029_Inner,XRISM2025_A2029_Outer} and is often regarded as one of the most relaxed clusters in the nearby Universe. Despite this, its core ($\lesssim200\kpc$) exhibits similarities to Perseus, including a prominent sloshing spiral \citep[e.g.,][]{Paterno-Mahler2013,Watson2026} and $\sim10-30\kpc$-scale radio lobes inflated by the central supermassive black hole as revealed by recent high-resolution LOFAR observations \citep{Timmerman2022}. The sharp and compact sloshing spiral arm suggests the merger likely occurred recently and the merger plane is not far from the plane of the sky. A southern infalling group located at a projected distance of $\sim2.5\Mpc$ from the cluster is a promising candidate for driving the merger. The group's asymmetric X-ray morphology (see \citealt{Mirakhor2022}) resembles a slingshot tail produced by gas stripping, which is commonly observed in off-axis minor mergers as the subhalo approaches apocenter \citep{Sheardown2019}. One of the most representative examples is the Coma-NGC~4839 system \citep[e.g.,][]{Lyskova2019,Churazov2021}. Such a stripped tail provides strong constraints on the overall merger configuration of the system.

Given the system's relatively high redshift and the moderate angular resolution of XRISM/Resolve, detailed modeling of the sloshing kinematics is essential for characterizing the merger contribution to the measured velocity dispersion. To this end, we perform tailored merger simulations of A2029 and the infalling group. Similar simulations have been carried out for multiple clusters to interpret XRISM velocity measurements (e.g., A1914, A3571, and Coma; see \citealt{Heinrich2025,Aihara2026,Zhang2026_Coma}, where one can find our detailed numerical setups). Based on weak-lensing estimates \citep{Sohn2019}, the merger mass ratio is fixed at 10:1 in all our runs. The gas density radial profile of the main cluster is adopted from the best-fit model for A2029 presented by \citet{Vikhlinin2006}. That of the group follows a Burkert profile \citep{Burkert1995}, which does not significantly affect the velocity structure formed in the core of the main cluster due to the large impact parameter in the model. In our numerical experiments, we vary the impact parameter to adjust the initial angular momentum and identify the viewing angle that best matches the observations.

Fig.~\ref{fig:merger_mock_a2029} shows the numerical prediction based on the simulation with an initial impact parameter of $3\Mpc$ and a pairwise velocity of $700\kms$. The snapshot corresponds to $t\simeq1.6\Gyr$ after the group's primary pericentric passage. The viewing direction is inclined by $i=30^{\circ}$ with respect to the normal of the merger plane (hereafter referred to as the inclination angle). The simulated merging system captures the key observational features of A2029, including the reproduction of the winding sloshing arm in both morphology and spatial scale \citep{Watson2026}. The infalling group is approaching its apocenter with a relative LOS velocity of $\simeq80\kms$ with respect to the main cluster, which is slightly smaller than the velocity implied by the redshift difference between the BCG of A2029 and the group ($\simeq150\kms$; see \citealt{Sohn2019}). Enhanced velocity dispersions are predicted near the sloshing cold fronts, but these regions are difficult to spatially resolve with XRISM (see \citealt{Uchida2026} for a recent XRISM sub-FOV analysis). However, such velocity dispersion jumps have been identified in Perseus near its sloshing arms \citep{XRISM2026_Perseus}. High velocity dispersions (e.g., $\gtrsim300\kms$) are expected in the turbulent wake of the infalling group along its orbital trajectory, which is not detected even in the outermost XRISM pointing along the NE arm. This non-detection places a lower limit on the orbital pericenter, implying either a large impact parameter, a small inclination angle, or both. In addition, mergers with excessively large initial impact parameters (e.g., $5\Mpc$) cannot perturb the core of the main cluster sufficiently to reproduce the observed X-ray structures based on our numerical experiments. We note that, although different merger scenarios for A2029 were proposed in the literature \citep[e.g.,][]{Paterno-Mahler2013,Sohn2019,Watson2026}, we present the first merger simulation tailored to A2029 and reproduce its major X-ray structures.

\begin{figure}
\centering
\includegraphics[width=0.95\linewidth]{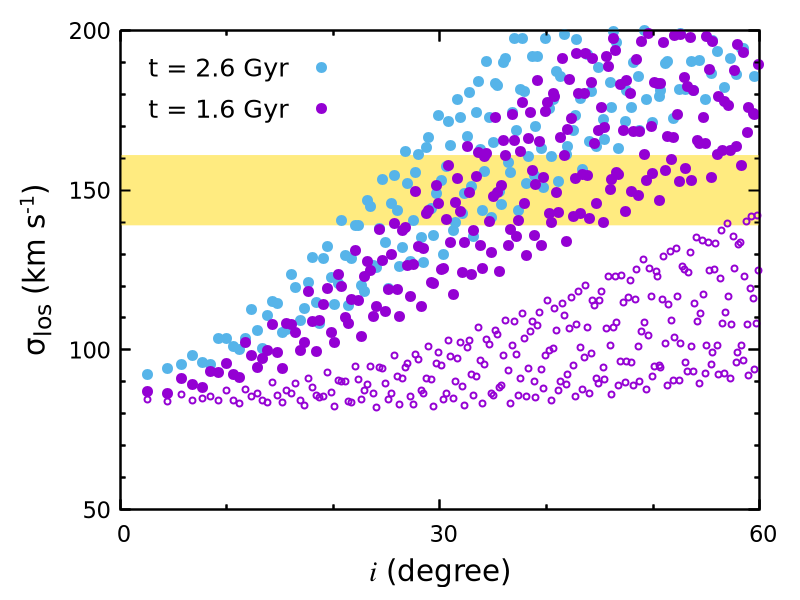}
\vspace{-5pt}
\caption{X-ray-weighted velocity dispersions (solid points) measured within a radius of $\simeq135\kpc$, approximately the XRISM/Resolve's FOV at A2029's redshift, along $500$ different sightlines in the $t=1.6$ and $2.6\Gyr$ snapshots, respectively. The sightlines are uniformly sampled over $4\pi$ steradians, with the inclination angle defined as the angle between the sightline and the normal to the merger plane. The horizontal yellow band indicates the XRISM measurement and its $1\sigma$ uncertainty. Open points show the velocity dispersion at $t=1.6\Gyr$ after excluding the bulk velocity contribution (see Section~\ref{sec:result:a2029} for details). The sloshing-driven bulk rotation of the gas is the major source of uncertainty in our model of $\sigma_{\rm los}$. }
\label{fig:merger_a2029_angle}
\end{figure}

\begin{figure}
\centering
\includegraphics[width=0.95\linewidth]{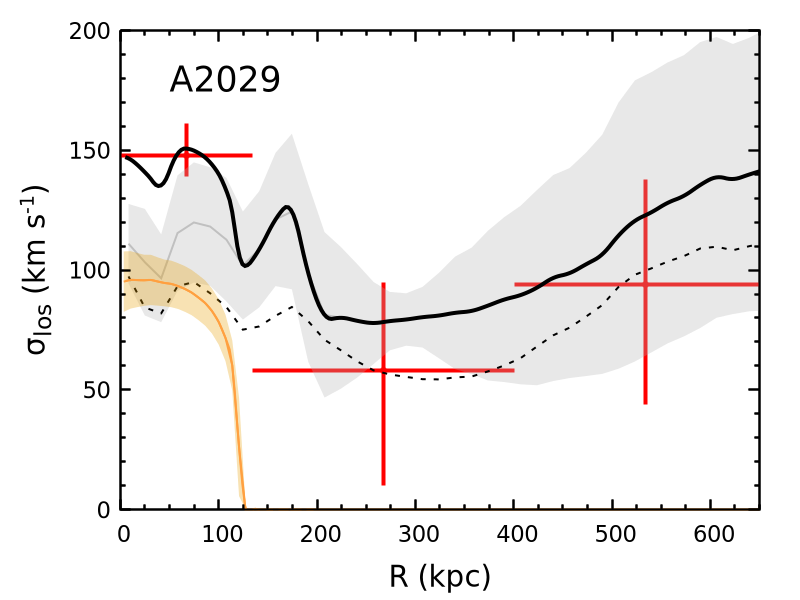}
\vspace{-5pt}
\caption{Same as Fig.~\ref{fig:perseus_model} but for A2029. The AGN feedback component is modeled with $\sigma_{\rm agn}=100\kms$, $\ell_{\rm inj}=20\kpc$, and $r_{\rm agn}=120\kpc$. The large-scale component (grey solid line and shaded region) is derived from the tailored merger simulation for A2029 at $t=1.6\Gyr$. The overlaid dashed line shows the predicted merger component with the bulk-motion contribution excluded, providing a better comparison with the measurements beyond $\simeq150\kpc$ along the NE arm \citep{XRISM2025_A2029_Outer}. Accounting for systematic uncertainties in the modeling, we can only place an upper limit on  $\sigma_{\rm agn}$, which nevertheless indicates that the AGN contribution is subdominant to that of the merger (see Section~\ref{sec:result:a2029}). }
\label{fig:a2029_model}
\end{figure}

\begin{figure*}
\centering
\includegraphics[width=0.95\linewidth]{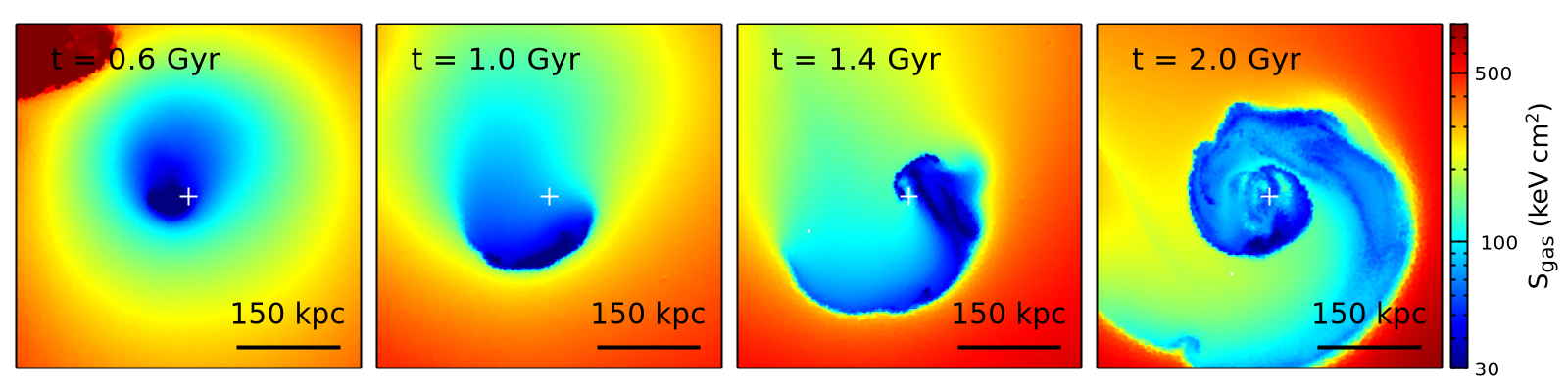}
\vspace{-5pt}
\caption{Gas entropy slices across the merger plane, illustrating the early evolution of atmospheric sloshing in our A2029 simulation. The low-entropy gas core ($\lesssim50\keV\cm^2$) is significantly displaced from the gravitational potential center (white cross) for $\sim0.8\Gyr$, increasing the central cooling time $t_{\rm cool}$ and the ratio of $t_{\rm cool}/t_{\rm ff}$. Our best-matching snapshot for A2029 is at $t\simeq1.6\Gyr$ (as shown in Fig.~\ref{fig:merger_mock_a2029}), approximately $0.2\Gyr$ after the low-entropy gas returns to the central region. This is broadly consistent with the lack of cold gas but the presence of newly launched radio jets in the system (see Section~\ref{sec:discussion:sample}). }
\label{fig:merger_a2029_slice}
\end{figure*}

To evaluate how the viewing angle and merger timing impact our velocity predictions in simulations, and hence the potential systematic uncertainties in our modeling, we compare velocity dispersions from the same region (white circle in Fig.~\ref{fig:merger_mock_a2029}) but along different sightlines and at different simulation snapshots, as shown in Fig.~\ref{fig:merger_a2029_angle}. We sample $500$ LOS directions uniformly over $4\pi$ steradians. The snapshot at $t=2.6\Gyr$ is chosen for comparison, at which point the merger-driven sloshing spiral has fully developed (see Fig.~\ref{fig:merger_a2029_slice}). The results show that more developed sloshing structures yield slightly higher $\sigma_{\rm los}$ values. Meanwhile, $\sigma_{\rm los}$ depends strongly on the inclination angle, which is largely due to the sloshing-driven rotation. The measured velocity dispersion can be written as
\be
\overline{\sigma_{\rm los}^2} = \Big<\sigma_{\rm los}(x,\,y)^2 + u_{\rm los}(x,\,y)^2\Big> -  \Big<u_{\rm los}(x,\,y)\Big>^2,
\ee
where $<>$ denotes the X-ray surface brightness weighted average over the sky region, and $\sigma_{\rm los}(x,\,y)$ and $u_{\rm los}(x,\,y)$ are X-ray-emissivity-weighted LOS velocity dispersion and bulk velocity at the projected sky coordinate $(x,\,y)$. The contribution from the term $\Big<u_{\rm los}(x,\,y)\Big>^2$ is small, but $\Big<u_{\rm los}(x,\,y)^2\Big>$ is sensitive to the inclination angle due to the quasi-axisymmetric bulk velocity distribution within the central region. In Fig.~\ref{fig:merger_a2029_angle}, we also show $\sqrt{\Big<\sigma_{\rm los}(x,\,y)^2\Big>}$ at $t=1.6\Gyr$ as open purple points, which exhibit a much weaker angle dependence. The difference between the open and solid points reflects the rotation contribution. Spatially resolving the coherent bulk velocity pattern within $R\lesssim150\kpc$ will provide a better constraint on the feedback component, $\sigma_{\rm agn}$ in particular. \citet{Uchida2026} recently reported XRISM sub-FOV measurements of A2029, revealing a central rotation pattern. Within the central pointing, we estimate their $\sqrt{\Big<u_{\rm los,sub}^2\Big>}$~\footnote{Calculated as the average over the subregions, weighted by their respective best-fit normalizations.} to be $\simeq100\kms$ and $60\kms$ with and without SSM correction, respectively. Our numerical prediction is closer to the value obtained without the SSM correction. Note that the limited azimuthal coverage of the outer cluster region may compromise the robustness of the SSM correction in A2029, given the significant large-scale velocity shear. Additional XRISM pointings with broader azimuthal coverage are critical for improving both the velocity measurements and merger modeling.

\begin{figure}
\centering
\includegraphics[width=0.95\linewidth]{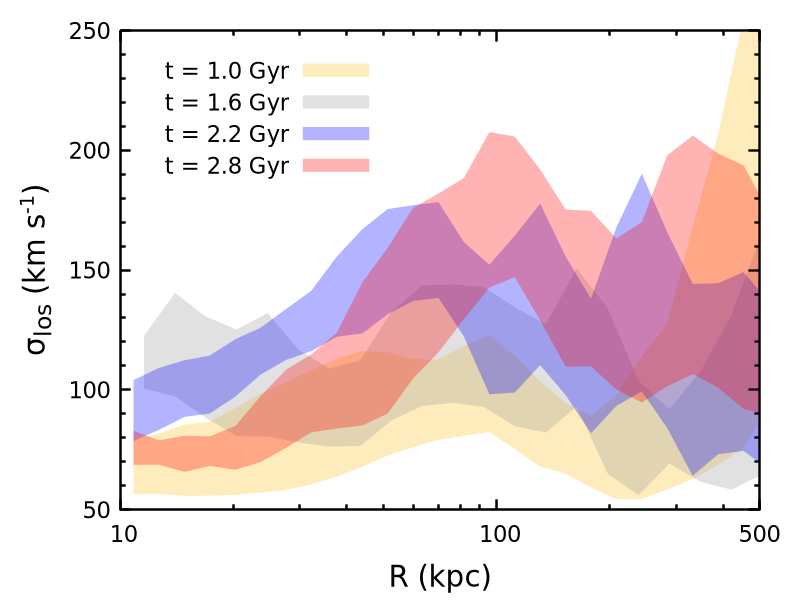}
\vspace{-5pt}
\caption{Evolution of the radial profile of the LOS velocity dispersion for the A2029 merger simulation, viewed along the direction adopted in the best-match model. The grey band shows the same data as in Fig.~\ref{fig:a2029_model}, but binned logarithmically in radius. The elevated $\sigma_{\rm los}$ within $\simeq40\kpc$ at $t=1.6\Gyr$ reflects the early stage of sloshing spiral development (see Section~\ref{sec:result:a2029}).}
\label{fig:merger_a2029_prof}
\end{figure}

Despite these uncertainties, we constrain the AGN feedback contribution in the core of A2029 based on the merger simulation. Fig.~\ref{fig:a2029_model} shows the double-driver decomposition of the velocity dispersion. The merger component is estimated from the azimuthally-averaged radial profile of the X-ray-weighted velocity dispersion derived from our best-matching snapshot, shown in Fig.~\ref{fig:merger_mock_a2029}. It is broadly consistent with the observational data beyond $R\sim150\kpc$. The overlaid dashed line shows the radial profile of $\sqrt{\Big<\sigma_{\rm los}(x,\,y)^2\Big>}$, which is in better agreement with the measurements along the NE arm alone. The predicted central velocity dispersion averaged within the XRISM FOV is $\simeq110\kms$, or $\simeq90\kms$ when the contribution from rotation is excluded. This relatively high merger-driven $\sigma_{\rm los}$ within $\simeq100\kpc$ in A2029, compared with Perseus, reflects the early stage of sloshing spiral development. Fig.~\ref{fig:merger_a2029_prof} shows the evolution of the radial profile of the LOS velocity dispersion. The elevated central $\sigma_{\rm los}\gtrsim100\kms$ persists for less than $1\Gyr$, after which the radial profile gradually evolves towards the shape modeled in Perseus (see Fig.~\ref{fig:perseus_model}).

The AGN feedback component is included to match the innermost XRISM measurement in Fig.~\ref{fig:a2029_model}. The transition radius between the two drivers is expected to be $\lesssim100$--$150\kpc$, assuming that the sloshing cold front plays a significant role in confining feedback-driven turbulence, as observed in Perseus along the NW direction. Our AGN feedback model adopts $r_{\rm agn}=120\kpc$ and $\sigma_{\rm agn}=100\kms$, corresponding to a $\sim40\%$ contribution from AGN feedback to the gas kinetic energy within $135\kpc$. The current XRISM measurements cannot place a lower limit on $r_{\rm agn}$ in A2029 (cf., A1795), and the feedback region may therefore be more radially concentrated.

We emphasize that systematic uncertainties in our modeling may affect the estimated AGN feedback contribution. A full velocity map of A2029 (e.g., extending to $\simeq500\kpc$) would provide stronger constraints on the large-scale bulk-motion pattern and, in turn, on the merger configuration and viewing direction. Other factors not captured by our idealized merger simulations may also affect the inferred velocity structure, such as the gravitational influence of the massive elliptical BCG, which may partially explain the more elongated sloshing structure observed along the N-S direction \citep{Watson2026}, close to the major axis of the BCG. In addition, pre-existing turbulent motions associated with historical merger events may further reduce our estimate of the AGN feedback contribution. Assuming a background turbulent velocity dispersion of $\simeq50$--$60\kms$ as in A1795, the inferred $\sigma_{\rm agn}$ would decrease to $\simeq80\kms$. Moreover, if instead we adopt the bulk velocity distribution measured with SSM correction from \citet{Uchida2026} as the reference model, we obtain $\sigma_{\rm agn}\sim0$. Our estimate of the AGN feedback contribution should therefore be regarded as an approximate upper limit.

\section{Discussions} \label{sec:discussions}

\subsection{Diverse turbulent driving scenarios} \label{sec:discussion:sample}

Among the three cool-core clusters discussed, Perseus and A1795 exhibit prominent kinematic signatures of AGN feedback with $\sigma_{\rm agn}\simeq120-150\kms$, indicating the presence of a kinematic sphere of influence ($r_{\rm agn}\sim50-100\kpc$) within which AGN feedback dominates the velocity field on scales smaller than $\ell_{\rm eff}$. Both systems show clear evidence of feedback activity in high-resolution X-ray and radio images, including AGN-driven bubbles and bubble-uplift features on scales of tens of kpc. They also host multiphase gas, including atomic and molecular phases traced by H$\alpha$ and CO \citep[e.g.,][]{Conselice2001,Crawford2005,Salome2008}, respectively. This suggests a plausible physical connection between ongoing AGN activity, particularly the episodes of AGN bubbles, and significant feedback-driven turbulence.

Buoyant bubbles generate turbulence directly in their wakes, while the uplifted low-entropy ICM drives large-scale circulation and triggers hydrodynamic instabilities. In both Perseus and A1795, the radial profile of $N_{\rm BV}$ is non-monotonic, peaking at $r\simeq50-60\kpc$ and thus forming a barrier to the radial propagation of the elongated bubble wakes and associated ICM entrainment from the cluster center. Eddy detachment naturally occurs near the $N_{\rm BV}$ peak \citep{Zhang2022}, which may help shape the flat entropy core of the ICM. It can also explain the $\sim40$--$50\kpc$ maximum lengths of the H$\alpha$ filaments observed in both Perseus and A1795 \citep{Conselice2001,Fabian2001}.

A2029 is in a somewhat different situation, with AGN feedback contributing less prominently to the velocity field within the core, despite its typical strong cool core and the presence of radio jets. A2029 has often been regarded as an outlier among cool-core clusters as it shows little to no cold gas or star formation \citep[e.g.,][]{McNamara2016,Olivares2023}. We speculate that these features may all be related to the recent off-axis minor merger. A similar picture has been suggested by \citet{Clarke2004}. In Fig.~\ref{fig:merger_a2029_slice}, we show the evolution of the gas entropy in a slice across the merger plane. The sloshing begins to displace the low-entropy ($\lesssim50\keV\cm^2$) central gas at $t\simeq0.6\Gyr$. With the maximum displacement reaching $\sim100\kpc$, this process reduces the efficiency of radiative cooling in the cluster core through gas expansion and redistribution. The pre-existing cold gas near the center might be also swept away at this stage. The low-entropy gas falls back later at $t\simeq1.4\Gyr$ in our simulation, on a timescale of about half the buoyancy period, i.e., $\sim\pi/N_{\rm BV}\simeq0.5$--$1\Gyr$. After that, the ratio of the cooling to free-fall timescales, $t_{\rm cool}/t_{\rm ff}$, decreases again toward the cluster center, potentially enabling thermal instabilities to develop \citep[see, e.g.,][and references therein]{Voit2015}. Similar gas behavior has been found in previous numerical studies \citep[e.g.,][]{Gomez2002,ZuHone2010}, demonstrating that sloshing can delay a build-up of cool gas. Our best-matching snapshot is $\sim0.2\Gyr$ after the return of the low-entropy gas, a timescale shorter than or comparable to the minimum $t_{\rm cool}$ of the system. This may partially explain why cold gas has not yet formed efficiently in A2029, although it is expected to occur shortly thereafter (see \citealt{Prasad2024} for an alternative interpretation). Observationally, the X-ray-bright spiral edge coils towards the BCG center, hinting that the developing cooling flow is fueling the central supermassive black hole. The inflated radio lobes are oriented perpendicular to the spiral structure near their base \citep{Clarke2004,Paterno-Mahler2013}, and their lifetime is $\simeq0.1\Gyr$ \citep{Birzan2004,Timmerman2022}, consistent with our scenario.

\subsection{A unified picture: an evolutionary view of the double cascade} \label{sec:discussion:picture}

\begin{figure}
\centering
\includegraphics[width=0.95\linewidth]{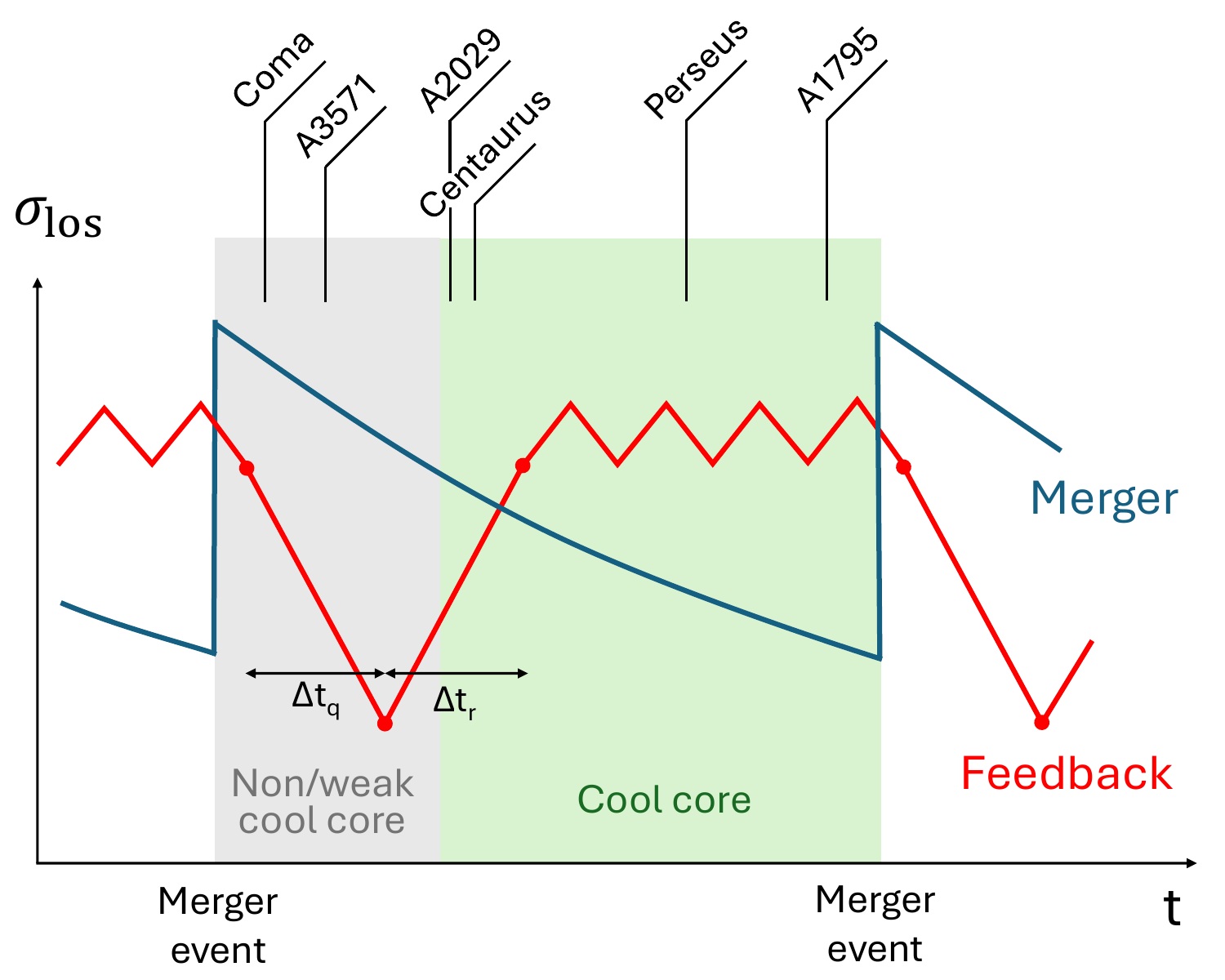}
\vspace{-0pt}
\caption{A schematic illustrating the relative contributions of merger-driven and AGN feedback-driven turbulence to the velocity dispersion measured by XRISM in cluster central regions. Mergers inject substantial kinetic energy during core passage, which gradually dissipates (blue curve). The characteristic timescale over which a cluster is affected by a significant merger is a few Gyr, depending on its merger history. The red curve indicates the AGN feedback contribution, which may be quenched or significantly suppressed by the violent redistribution of the gas during the merger, but is eventually restored as the atmosphere resettles and cools. $\Delta t_{\rm q}$ and $\Delta t_{\rm r}$ denote the quenching and turbulence-refilling timescales, respectively. The sketch is not drawn to scale. A quasi-steady state of AGN feedback is likely established in relaxed clusters. The zig-zags are intended to illustrate the spatial and temporal variations in AGN feedback. We mark six galaxy clusters with deep XRISM observations in the sketch, including the two clusters analyzed in detail in this paper as well as Perseus \citep{XRISM2026_Perseus,Zhang2026}, A3571 \citep{Aihara2026,McCall2026}, Centaurus \citep{XRISM2025_Centaurus}, and Coma \citep{XRISM2025_Coma,Gatuzz2026}. }
\label{fig:sketch}
\end{figure}

To interpret the diverse physical origins of turbulent motions in cluster cores, we propose a unified picture in which clusters undergo similar evolutionary processes but are observed at different dynamical stages. In addition to the clusters analyzed in this paper, we include several other systems recently observed by XRISM in our discussion.

Fig.~\ref{fig:sketch} illustrates this picture, which consists of two main ingredients. First, energetic mergers drive the transition from cluster cool-core to non-cool-core/weak-cool-core states. In this process, mergers can quench or suppress AGN feedback by displacing, heating, and mixing the low-entropy atmosphere that is deeply confined by the gravitational potential, as demonstrated in the A2029 case (see \citealt{ZuHone2010,Chadayammuri2021} for more systematic theoretical studies). Second, the gas entropy gradient is re-established shortly after the merger, enabling more efficient formation of cold gas and eventually fueling the central black hole. The AGN feedback cycle is then recovered. The refilling of feedback-driven turbulence within the cluster cooling radius $r_{\rm cool}$, however, occurs on a longer timescale, introducing a delay of roughly the bubble-crossing time across $r_{\rm cool}$ (see more discussion in Section~\ref{sec:discussion:heating}). Meanwhile, the merger-driven large-scale motions cascade and gradually dissipate over a few dynamical timescales. Eventually, feedback-driven motions take over within the inner core and establish a kinematic sphere of influence, giving rise to the apparent double-cascade pattern observed in Perseus and A1795. A2029 may represent a transitional stage in which the AGN feedback has been re-ignited, but feedback turbulence has not yet overtaken the residual merger-driven motions. A similar case is the Centaurus cluster, whose motions are dominated by sloshing based on the recent XRISM measurements \citep{XRISM2025_Centaurus}, while it also exhibits clear signatures of AGN feedback, including X-ray bubbles and coexistence of multiphase gas. The latter suggests that either Centaurus is at a more advanced evolutionary stage than A2029 or the merger was insufficient to fully quench the feedback activity. Yet another possibility is that AGN feedback makes a substantial contribution to the measured velocity dispersion in the central region of Centaurus, but separating it from the merger component requires more spatially extended velocity mapping and detailed numerical modeling.

A3571 provides another interesting comparison. It is a weak cool-core system that exhibits clear sloshing structures but no unambiguous evidence of AGN feedback (see \citealt{McCall2026} and references therein). \citet{Aihara2026} argued that the system is in an early stage of sloshing development, with the dense core displaced from the center of the gravitational potential but without a spiral cold front having yet formed, roughly corresponding to the stages shown in the first two panels of Fig.~\ref{fig:merger_a2029_slice}. This provides a crucial observational counterpart to our picture, in which merger-driven sloshing disrupts the feedback activities. It may indeed represent a precursor to A2029, roughly $\sim0.5-1\Gyr$ earlier in its evolutionary sequence.

In this discussion, we primarily focus on significant minor mergers with mass ratios ranging from a few to a few tens. Such mergers occur at a rate of approximately a few events per unit redshift \citep[e.g.,][]{Fakhouri2010}. Major mergers play a similar role in quenching feedback, but they often involve nearly head-on, highly energetic collisions that can completely disrupt the cool core. The Coma cluster is such an example with evidence for a broad, flat core and the presence of two BCGs \citep[e.g.,][]{Briel1992}. However, \citet{Zhang2026_Coma} recently suggested that its recent minor merger with the galaxy group NGC~4839 is sufficient to explain the velocity structure measured by XRISM \citep{{XRISM2025_Coma,Gatuzz2026}}. In this sense, Coma is similar to A3571, but its AGN feedback activity may be more difficult to re-establish than in weak cool-core systems. Nevertheless, we include Coma in Fig.~\ref{fig:sketch}.

We acknowledge that our sketch likely oversimplifies the physical processes governing the transition between cool cores and non-cool cores \citep[e.g.,][]{Burns2008,Rossetti2010,Lehle2025} and the interaction between AGN feedback and mergers \citep[e.g.,][]{Chadayammuri2021,Chen2026}. Despite these simplifications, our results underscore the importance of investigating individual systems in detail when comparing and interpreting XRISM kinematic measurements across clusters. Similar velocity dispersions may arise from fundamentally different physical mechanisms. It is also worth noting that velocity mapping is essential for distinguishing the contributions from AGN feedback and mergers on different spatial scales. All systems shown in Fig.~\ref{fig:sketch}, except Centaurus,\footnote{An off-axis pointing of Centaurus has also been obtained but the corresponding measurement has not yet been published.} have been observed with multiple XRISM pointings.

\subsection{Turbulent heating} \label{sec:discussion:heating}

The role of turbulent dissipation in offsetting the ICM radiative cooling remains under debate. The turbulent heating rate per volume can generally be estimated as a function of the turbulent 3D velocity amplitude $u_{\rm k}$ and its corresponding scale $\ell_{\rm k}$, i.e.,
\be
q_{\rm heat} = \eta_0\frac{\rho_{\rm gas}u_{\rm k}^3}{\ell_{\rm k}},
\label{eq:qheat}
\ee
where $\eta_0$ is a constant coefficient and $\rho_{\rm gas}$ is the gas mass density. While XRISM enables direct measurements of gas velocities, the relevant spatial scales remain poorly constrained and depend on the mechanism driving the turbulence.

\begin{figure}
\centering
\includegraphics[width=0.95\linewidth]{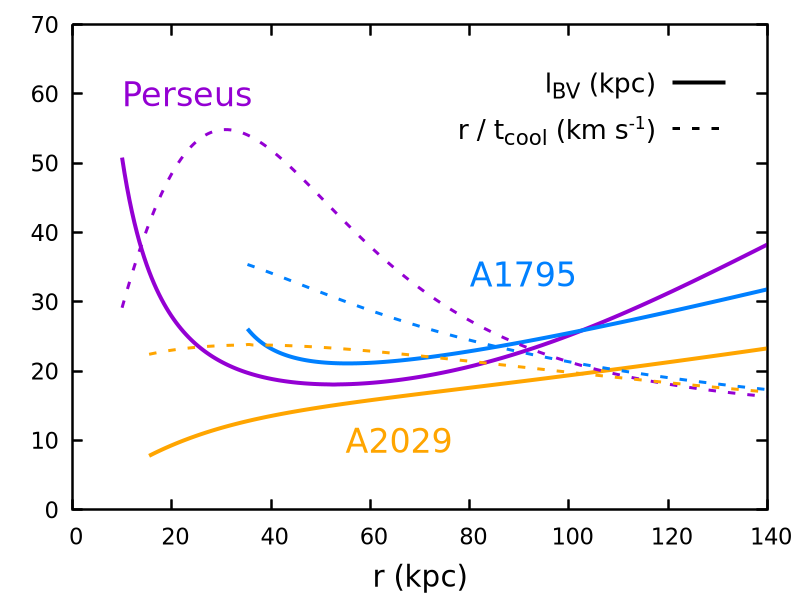}
\vspace{-5pt}
\caption{Radial profiles of the Ozmidov length scale $\ell_{\rm BV}$ (solid lines) and the characteristic velocity $r/t_{\rm cool}$ (dashed lines) for Perseus, A1795, and A2029. Rising bubbles drive turbulence through their interaction with the cluster atmosphere, with the maximum radial scale of the turbulence limited by the local $\ell_{\rm BV}$. The velocity $r/t_{\rm cool}$ represents the minimum energy-transport velocity required to keep pace with radiative cooling and is generally an order of magnitude lower than the bubble terminal velocity in the systems (see Section~\ref{sec:discussion:heating}).
}
\label{fig:profiles}
\end{figure}

Merger-driven turbulence is generated on large scales but is subsequently regulated by gravitational stratification and the spherical geometry of the potential well. As the cluster relaxes following the merger, the turbulence decays and gradually becomes confined to radial layers, resulting in anisotropic velocity fields whose properties depend on the eddy Froude number \citep[e.g.,][]{Shi2019,Wang2023,Lau2026}. Assuming that the largest available eddy size at each radius is $\simeq r$ and that these eddies contain most of the turbulent kinetic energy \citep[e.g.,][]{Shi2014}, the heating rate is proportional to $\sigma_{3D}(r)^3/r$, where $\sigma_{3D}(r)$ is the 3D gas velocity dispersion. To connect this with the velocities measured by XRISM, $\sigma_{\rm los}$, the X-ray emissivity distribution along the LOS needs to be taken into account. The heating rate therefore scales as $\sigma_{\rm los}^3/\ell_{\rm eff}$, since the turbulence injection scale is generally larger than $\ell_{\rm eff}$ for mergers, with an additional factor that depends on the degree of turbulence anisotropy (e.g., a factor of $3^{3/2}$ for isotropic turbulence). In cool-core clusters, merger-driven turbulence alone is generally insufficient to balance radiative cooling losses within the central dense region, although it may provide sufficient heating to offset cooling in weak cool-core systems (e.g., A3571 and A1413; see \citealt{McCall2026,Gormez2026}).

At a similar velocity dispersion amplitude (e.g., $\simeq100-150\kms$), feedback-driven turbulence can provide a higher heating rate because it is generated on relatively smaller scales \citep{XRISM2026_Perseus}. For turbulence directly produced by rising bubbles, the dominant scale is set by the bubble size, the Ozmidov scale $\ell_{\rm BV}$, or both. In Fig.~\ref{fig:profiles}, we show the radial profiles of $\ell_{\rm BV}$ for our clusters (solid lines), which are estimated from the best-fit deprojected gas density and temperature profiles reported by \citet{Tang2017} for Perseus and by \citet{Vikhlinin2006} for A1795 and A2029. Both Perseus and A1795 exhibit a weakly stratified regime in their inner regions, defined by $r<\ell_{\rm BV}(r)$, within which the maximum turbulent eddy size is limited primarily by the available volume, reaching $\simeq25$ and $30\kpc$ in Perseus and A1795, respectively. At larger radii, the Ozmidov scale sets the maximum bubble eddy size in the radial direction.

\begin{figure}
\centering
\includegraphics[width=0.95\linewidth]{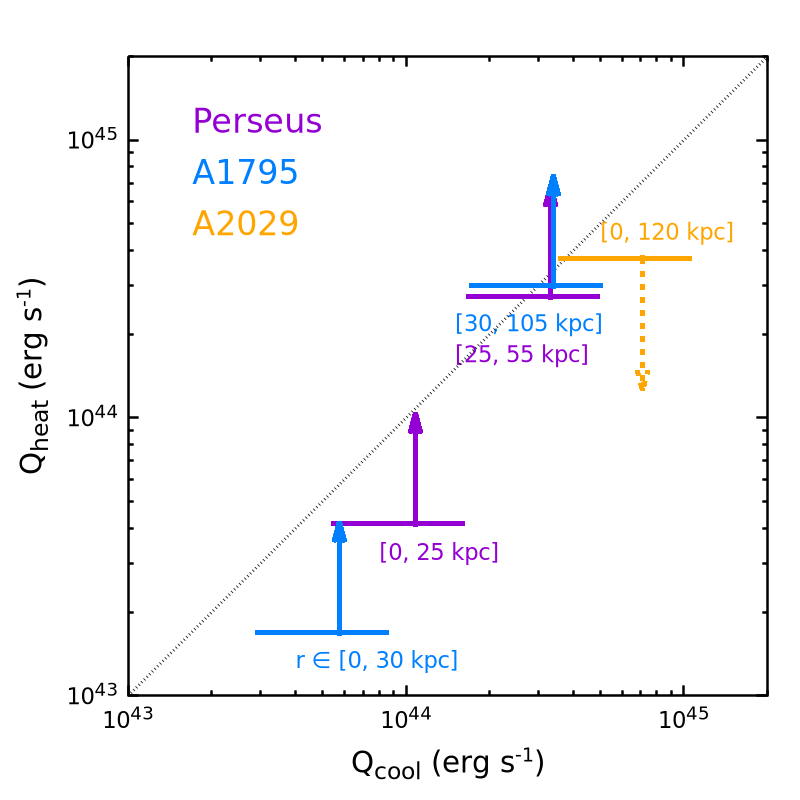}
\vspace{-5pt}
\caption{Comparison of the AGN feedback-driven turbulent heating rate with the radiative cooling rate within $r_{\rm agn}$. For Perseus and A1795, we separately compare the rates in the inner weakly stratified and outer regions, with their radial ranges marked in the figure. The estimated heating rates should be regarded as lower limits because the turbulent distribution is likely non-uniform and the relevant turbulent scales may be smaller than those adopted. We conservatively assign a $\pm50\%$ uncertainty to the cooling rates following the argument in \citet{XRISM2026_Perseus}. For A2029, we obtain only an upper limit on $\sigma_{\rm agn}$ and hence on the corresponding heating rate. In Perseus and A1795, the turbulent heating rates are broadly comparable to the cooling rates, highlighting the potentially significant role of turbulence in AGN feedback heating (see Section~\ref{sec:discussion:heating}).  }
\label{fig:heating}
\end{figure}

Based on this picture, we estimate the turbulent heating rate due to AGN feedback using the coefficient from \citet{Zhuravleva2014} for isotropic Kolmogorov turbulence ($\eta_0\simeq1$) and $u_{\rm k}=\sqrt{3}\sigma_{\rm agn}$ in Eq.~\ref{eq:qheat}, and compare it with the radiative cooling rate, shown in Fig.~\ref{fig:heating}. Note that the adopted characteristic turbulent scale corresponds to a Froude number of unity, at which the turbulence is not expected to deviate significantly from isotropy. For Perseus and A1795, we separately estimate the rates in the inner weakly-stratified region and the outer part ($<r_{\rm agn}$), adopting the corresponding maximum eddy size of $\ell_{\rm k}=25\kpc$ and $20\kpc$ for Perseus, and $\ell_{\rm k}=30\kpc$ and $25\kpc$ for A1795, respectively, in the radial direction. These estimates likely represent lower limits for the following two reasons. First, the turbulent scales are not directly measurable with XRISM. The actual injection scales can be smaller, particularly in the inner regions, based on the morphology of the X-ray and radio bubbles (see, e.g., \citealt{Timmerman2022}). This is further supported by the flattening of the velocity structure function of the H$\alpha$ filaments at $\sim10\kpc$ in both systems \citep{Li2020,Tamhane2023}. Second, we assume a spatially uniform turbulent distribution to constrain the volume-averaged $\sigma_{\rm agn}$ for simplicity. Introducing a volume-filling factor $f\ (\leq1)$ would increase the inferred heating rate as $q_{\rm heat}\propto f^{-1/2}$ for a fixed volume-averaged velocity amplitude $\sigma_{\rm agn}$.

Despite the large uncertainties, the estimated heating rates are broadly comparable with the cooling rates in Perseus and A1795, highlighting the potentially significant role of turbulence in AGN feedback \citep{Zhuravleva2014,XRISM2026_Perseus}. A1795 is dynamically more relaxed than Perseus, with a weaker impact of merger activity on the AGN jets. The ratio of turbulent heating to radiative cooling is interestingly similar in the two systems, hinting at an established quasi-steady state in which the turbulent heating is regulated by feedback.

We emphasize that perfect cooling-heating balance, either spatially or temporally, is not required for radio-mode AGN feedback to prevent the ICM from catastrophic overcooling over the cooling timescale. Turbulence can locally overheat the ICM, particularly around rising bubbles and along their trajectories. The excess energy can then be redistributed through convection and mixing. In our estimates, the turbulent turnover timescale is $\sim\ell_{\rm k}/\sigma_{\rm k}\lesssim1/N_{\rm BV}$, shorter than the ICM cooling timescale, allowing efficient dissipation of turbulent energy. Meanwhile, for turbulence to play an important role in AGN feedback, its energy must be replenished, on average, faster than it is lost to radiative cooling. The characteristic velocity $r/t_{\rm cool}$ represents the minimum energy-transport speed required to keep pace with radiative cooling at a given radius, whether the transport is radial or azimuthal, as indicated by the dashed lines in Fig.~\ref{fig:profiles}. This velocity is generally small, being a factor of $\sim3-5$ lower than $\sigma_{\rm agn}$ and at least a factor of $\sim10$ lower than the bubble terminal velocity in the ICM \citep{Churazov2001,Zhang2018}. Bubbles can therefore efficiently deposit kinetic energy throughout the cooling radius on timescales much shorter than $t_{\rm cool}$. Adopting a diffusion coefficient of $D\sim\sigma_{\rm agn}\ell_{\rm BV}\simeq10^{30}\,{\rm cm^2\,s^{-1}}$, admittedly only an order-of-magnitude estimate, gives a mixing timescale across a distance $L$ of $t_{\rm mix}\simeq30\Myr\,(L/10\kpc)^2$. This timescale remains shorter than the cooling time throughout the radial range where AGN feedback is required to prevent significant overcooling ($\lesssim100\kpc$).

For A2029, we obtain only an upper limit on $\sigma_{\rm agn}$. Although $\ell_{\rm k}=15\kpc$ is also an upper limit following the same argument as for the other clusters, we conservatively regard the resulting heating rate as an upper limit given that $q_{\rm heat}\propto\sigma_{\rm agn}^3$. It is therefore possible that turbulent heating, even including the merger contribution, is insufficient to offset the radiative cooling in A2029 at the present epoch \citep[see also][]{Uchida2026}. This is broadly consistent with the X-ray surface-brightness fluctuation analysis of \citet{Zhuravleva2018} and supports the picture described in Section~\ref{sec:discussion:sample}: the atmosphere of A2029 is currently underheated, but for a duration shorter than the minimum $t_{\rm cool}$ of the system. Thus, substantial amounts of cold gas have not yet formed, while the cooling imbalance may have already triggered new AGN activity.

Internal gravity waves are inevitably excited during AGN feedback processes through the interaction between rising bubbles and the stratified ICM \citep[e.g.,][]{Zhang2018}. Their energy partition, power spectrum, and spatial distribution, however, remain poorly constrained in the cluster atmosphere. On the one hand, they enable efficient energy transport to the far field, particularly in the azimuthal direction; on the other hand, their presence may bias estimates of the turbulent heating rate, as their dissipation differs substantially from the Kolmogorov picture \citep[see][for a review]{Staquet2002}. When viscosity is low, the energy of large-scale gravity waves generally needs to be transferred to smaller scales through nonlinear wave interactions, wave breaking, or turbulence before it dissipates \citep[e.g.,][]{Muller1986,Sarkar2017}. The form of Eq.~\ref{eq:qheat} for the heating rate may still hold, but the coefficient and characteristic length scale depend on the specific nonlinear mechanism.

Current XRISM measurements cannot yet constrain the degree of spatial uniformity of the turbulence distribution because of the limited spatial resolution. In A1795, however, we tentatively find a uniform velocity distribution across the four quadrants on $\simeq100\kpc$ scales within the $1\sigma$ statistical uncertainties (see Fig.~\ref{fig:a1795_params}). In Perseus, \citet{XRISM2026_Perseus} reported $\sim1-2\sigma$ variations in velocity dispersion on $\simeq30\kpc$ scales. These results nevertheless suggest that turbulent kinetic energy may be efficiently replenished on the scales comparable to the cooling radius. A larger sample of cool-core clusters, together with detailed velocity mapping of nearby systems in the future, will provide further statistical insight into the spatial distribution of turbulent energy in the ICM.

\section{Conclusions} \label{sec:conclusions}

The double-cascade scenario was recently proposed for the Perseus cluster as a framework for explaining the complex core velocity field revealed by XRISM observations \citep{XRISM2026_Perseus}, in which AGN feedback and mergers drive turbulence predominantly at different radii and on distinct spatial scales. An important question is whether this picture is universal among cool-core clusters. In this work, we revisit recent XRISM observations of the cool-core clusters A1795 and A2029 to search for further evidence of this phenomenon. Our main findings are summarized below.
\begin{itemize}
  \item A1795 provides another clear example of the double-cascade picture after Perseus, with AGN feedback making a dominant contribution to the velocity-dispersion distribution within $\simeq100\kpc$. The motions are likely generated on relatively small scales ($\lesssim\ell_{\rm eff}$), consistent with the lack of significant bulk motions in the cluster core. This extended feedback-dominated kinematic region, together with the elongated H$\alpha$ and radio structures, suggests that rising bubbles may play a significant role in driving the motions. Meanwhile, the flat radial profile of the velocity dispersion disfavors sound waves as the dominant source of the velocity field.
  \item Under an assumption that the velocity field is dominated by AGN bubbles rising predominantly in the radial direction, we estimate the turbulent heating rate and find it to be comparable to the radiative cooling rate in A1795, similar to that found in Perseus. Despite the large uncertainties, this result highlights the potentially significant role of turbulence in AGN feedback.
  \item In contrast to A1795 and Perseus, AGN feedback makes only a subdominant contribution to the velocity field in A2029. We perform a tailored merger simulation for the system and find that it experienced an off-axis minor merger $\simeq1.6\Gyr$ ago with the group now located $\sim2\Mpc$ to the south, which perturbed the cluster atmosphere and displaced the central low-entropy gas. A2029 is now at an interesting dynamical stage: the displaced gas has just returned to the cluster center within $\sim0.2\Gyr$, a timescale shorter than the minimum cooling timescale of the system, potentially reigniting the AGN activity. This is consistent with the newly formed AGN radio lobes in the BCG and may partially explain the lack of cold gas (e.g., H$\alpha$) in this cool-core system.
  \item We propose a unified picture for interpreting the velocity fields of weak-cool-core and cool-core clusters recently observed by XRISM (see Fig.~\ref{fig:sketch}). In this picture, galaxy clusters undergo similar evolutionary processes but are observed at different dynamical stages. Significant minor mergers drive the transition from cool-core to non/weak-cool-core systems and quench or suppress AGN feedback activity. During and shortly after the merger, merger-driven large-scale motions dominate the velocity field, as in A3571. As the cluster relaxes and the merger-driven turbulence decays, feedback-driven turbulence re-emerges in the central region, as observed in Perseus and A1795, producing a clear double-cascade kinematic pattern. A2029 and Centaurus likely represent transitional stages in this evolutionary sequence.
\end{itemize}

Overall, our results demonstrate that AGN feedback-driven and merger-driven turbulence commonly coexist in cool-core clusters. Their spatial distribution and relative contributions to the velocity dispersion, however, depend on the merger history and evolutionary stage of individual systems. This complexity may also complicate the interpretation of radio mini-halos if turbulent re-acceleration is their dominant formation mechanism. We emphasize the importance of detailed studies of individual clusters to interpret the XRISM kinematic measurements and highlight the challenge of disentangling turbulent components arising from different physical processes. Spatially mapping the velocity field, together with detailed dynamical modeling, is essential for characterizing these components and understanding their underlying physics.

\begin{acknowledgements}

CZ and MZ acknowledge the support of the Czech Science Foundation (GACR) Junior Star grant no. GM24-10599M.
IZ acknowledges partial support by NASA grant 80NSSC18K1684 and the Alfred P. Sloan Foundation through the Sloan Research Fellowship. IK was supported by the Simons Foundation via the Simons Investigator Award to A. A. Schekochihin.
The simulations presented in this paper were carried out using the Midway computing cluster provided by the University of Chicago Research Computing Center.

\end{acknowledgements}

%
%

\begin{appendix} 
\onecolumn

\section{Turbulent simulations in a stratified atmosphere} \label{sec:appendix:sim}

\begin{figure*}[!h]
\centering
\includegraphics[width=1.\linewidth]{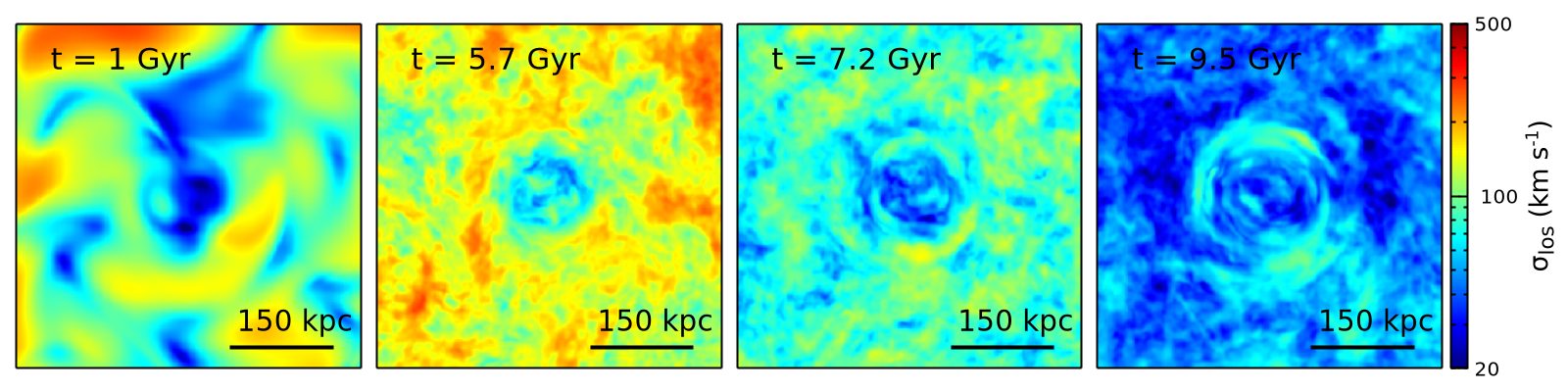}
\vspace{-10pt}
\caption{Evolution of the X-ray-weighted velocity dispersion distribution in our turbulent simulation. The turbulence-driving force was switched off at $t=5\Gyr$ to allow the turbulence to decay in the stratified cluster atmosphere. The three rightmost panels show the snapshots from which we extract the velocity dispersion radial profiles for comparison with Perseus ($t=5.7\Gyr$; see Fig.~\ref{fig:perseus_model}) and A1795 ($t=7.2$ and $9.5\Gyr$; see Fig.~\ref{fig:a1795_model}).}
\label{fig:turb_sim_perseus}
\end{figure*}

We adopt the 3D hydrodynamical turbulence simulation presented in \citet{XRISM2026_Perseus} to model the large-scale, merger-driven turbulence background in Perseus and A1795. Here, we summarize the key simulation settings; further numerical details are provided in the Supplementary Information of \citet{XRISM2026_Perseus}. The turbulence was driven on a scale of $\simeq500\kpc$ within a static, Perseus-like gravitational potential using a stochastic forcing field generated according to an Ornstein-Uhlenbeck process \citep[e.g.,][]{Federrath2010}. The driving force was switched off at $t=5\Gyr$, after which the turbulence was allowed to decay freely. Fig.~\ref{fig:turb_sim_perseus} shows the time evolution of the X-ray-weighted velocity dispersion distribution in the simulation. The projected velocity dispersion appears lower in the central core because of the smaller effective length, effectively reflecting the velocity power at smaller scales. The azimuthally-elongated structures in the maps arise from stratified turbulence and/or internal gravity waves.

\end{appendix}

\end{document}